# Direct identification of Mott Hubbard band pattern beyond charge density wave superlattice in monolayer 1T-NbSe$_2$

Liwei Liu[1*], Han Yang[1], Yuting Huang[2], Xuan Song[1], Quanzhen Zhang[1], Zeping Huang[1], Yanhui Hou[1], Yaoyao Chen[1], Ziqiang Xu[1], Teng Zhang[1], Xu Wu[1], Jiatao Sun[1], Yuan Huang[1,3], Fawei Zheng[4], Xianbin Li[2], Yugui Yao[4], Hong-Jun Gao[3], and Yeliang Wang[1]

[1]School of Information and Electronics, MIIT Key Laboratory for Low-Dimensional Quantum Structure and Devices, Beijing Institute of Technology, Beijing 100081, China.

[2]State Key Laboratory of Integrated Optoelectronics, College of Electronic Science and Engineering, Jilin University, Changchun 130012, China.

[3]Institute of Physics, Chinese Academy of Sciences, Beijing 100190, China.

[4]Key Lab of Advanced Optoelectronic Quantum Architecture and Measurement (MOE) and School of Physics, Beijing Institute of Technology, Beijing 100081, China.

*email: liwei.liu@bit.edu.cn

**Understanding Mott insulators and charge density waves (CDW) is critical for both fundamental physics and future device applications. However, the relationship between these two phenomena remains unclear, particularly in systems close to two-dimensional (2D) limit. In this study, we utilize scanning tunneling microscopy/spectroscopy to investigate monolayer 1T-NbSe$_2$ to elucidate the energy of the Mott upper Hubbard band (UHB), and reveal that the spin-polarized UHB is spatially distributed away from the d$z^2$ orbital at the center of the CDW unit. Moreover, the UHB shows a √3 × √3 R30° periodicity in addition to the typically observed CDW pattern. Furthermore, a pattern similar to the CDW order is visible deep in the Mott gap, exhibiting CDW without contribution of the Mott Hubbard band. Based on these findings in monolayer 1T-NbSe$_2$, we provide novel insights into the relation between the correlated and collective electronic structures in monolayer 2D systems.**

## Introduction

Mott insulators in coexistence with charge density wave (CDW) patterns in layered materials such as 1T-TaS$_2$[1-11] is a long-term and extremely challenging research topic. They have been investigated for approximately 40 years owing to their fascinating multiple CDW states[2]. These types of materials are associated with insulator-to-metal changes and possess promising application potential in future nanodevices, such as ionic field-effect transistors[12], voltage oscillators[13], and ultrafast switches or memories[10,14-18]. Mott state and CDW order are believed to be correlated. For example, the CDW order can induce a substantial narrowing of the band width near the Fermi level, and with further on-site Coulomb repulsion, a Mott gap is opened[19]. Hitherto, significant effort has been expended in manipulating the CDW order to induce the collapse of the Mott insulator, e.g., by pressure application[2,20], chemical doping[21,22], interfacial structure tuning[23,24], interlayer stacking[25-28], and pulse injection of ultrafast laser[25,29], current[30,31], or voltage[5-7]. The main efforts have focused on bulk $TaS_2$ crystals. Recently, the realization of the two-dimensional (2D) form of $TaS_2$ and similar transition metal dichalcogenides (TMD)[8,26,32] has caused the CDW order or Mott insulator to be revisited, with the aim to investigate the underlying mechanism and manipulate the layer-dependent properties of these novel quantum materials.

As a cousin of 1T-$TaS_2$, 1T-$NbSe_2$ has rarely been reported experimentally either in bulk or in a monolayer, because its favorable form is the 2H phase rather than the 1T phase[33]. Recently, some studies have reported the local induction of a 1T patch using the tip pulse[34] or crash[35] on the surface of a bulk 2H-$NbSe_2$; furthermore, one study reported the molecular beam epitaxy (MBE) growth of small monolayer islands[36]. All these studies revealed the Star of David (SOD) pattern, which is a type of CDW superlattice; however, the relation between the Hubbard band of Mott states with the CDW pattern has not been investigated.

In this study, we used the MBE method to grow high-quality large 1T-$NbSe_2$ islands (a single island can be larger than 140 nm) on bilayer graphene (BLG) supported on a SiC substrate, which provides a good platform for studying the relationship between the Mott Hubbard band and the CDW pattern with a high spatial and energy resolution. In fact, we observed three important features that have not been reported to date in 2D TMDs. First, the Mott upper Hubbard band (UHB) was distributed away from the center of the SOD. Second, the UHB exhibited a local $\sqrt{3} \times \sqrt{3}$ R30° periodicity related to the typically observed CDW pattern. Furthermore, a pattern similar to the CDW superlattice was visible in the Mott gap, showing a separation of the CDW from the Hubbard band. These observations provide some perspectives into the relation between Mott insulator bands and CDW in 2D quantum materials.

**Results**

**Star of David CDW pattern of ML 1T-$NbSe_2$.** Figure 1a shows a large-scale scanning tunneling microscopy (STM) image with monolayer (ML) $NbSe_2$ islands grown on BLG/SiC(0001). The 1T or 2H islands can be identified from the difference in the CDW superstructures, as the 1T phase has a larger modulation in the CDW pattern than the 2H phase (see Supplementary Fig. 1). Next, we focus on the ML1T-$NbSe_2$ islands. The center in Fig. 1a shows an ML 1T-island with a truncated triangular shape, with an edge length of ~120 nm. A schematic model of this ML 1T-$NbSe_2$ island is shown in Fig. 1b. The atomic resolution image of the island shows a clear CDW pattern (in Fig. 1c, the inset is the fast Fourier transform (FFT) of this pattern), and a zoomed-in image (Fig. 1d) shows the basis vectors of both the atomic lattice in the topmost layer and the CDW pattern ($a_1$ and $a_2$; $b_1$ and $b_2$, respectively). The CDW order with a superstructure of $\sqrt{13} \times \sqrt{13}$ R13.9° to the topmost layer can be well distinguished. This type of CDW pattern is often known as the SOD pattern, reminiscent of those in 1T-TaS$_2$[37,38] and 1T-TaSe$_2$[8,32].

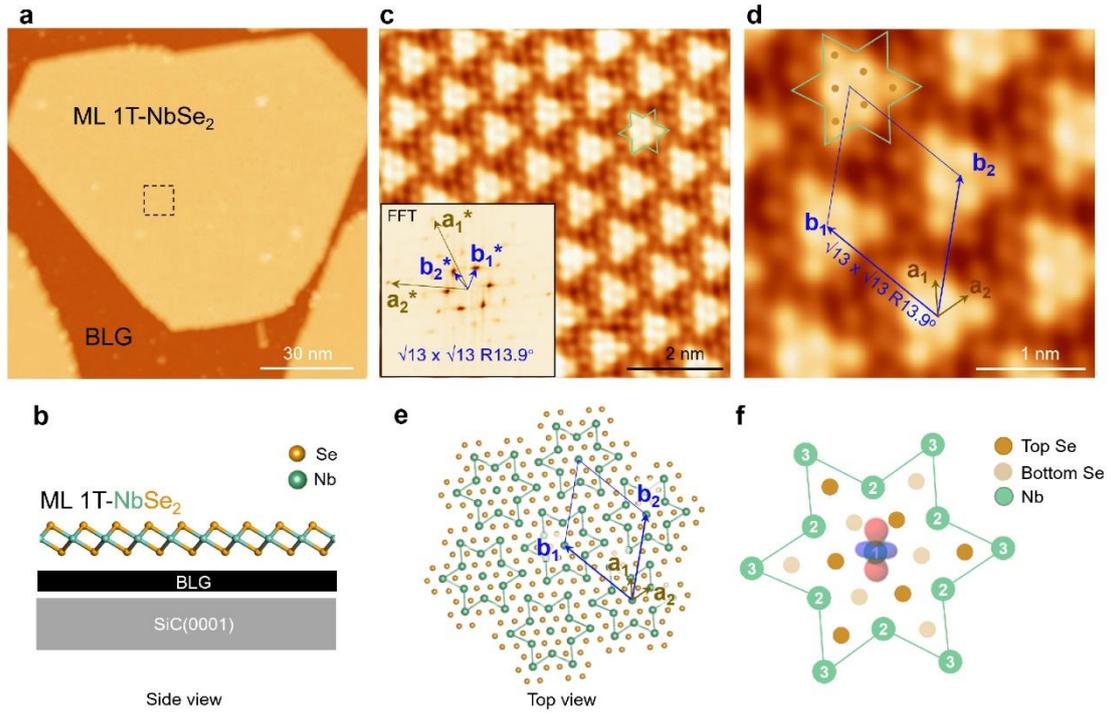

**Fig. 1 Large-scale and high-quality monolayer 1T NbSe$_2$ with "Star of David" (SOD) CDW pattern. a** Large-scale STM image of MBE-grown NbSe$_2$ islands on BLG/SiC(0001), showing ML 1T-NbSe$_2$ island with edge length 120 nm at the center. **b** Side view of schematic atomic model of ML 1T-NbSe$_2$ on BLG/SiC(0001) substrate. **c** Zoomed-in image of dashed black square in **a** showing both CDW pattern and atomic resolution of ML 1T-NbSe$_2$ surface. Inset shows FFT image. **d** Zoomed-in image showing basis vector of CDW unit cell (with unit vectors b$_1$ and b$_2$) and atomic lattice of Top-Se layer (with unit vectors a$_1$ and a$_2$); the SOD (green star) was imaged as six top Se atoms (brown dots). **e** Top view of atomic model of NbSe$_2$, showing SOD pattern and rhombus unit cell. **f** SOD model showing 13 Nb atoms at three different equivalent sites with six-fold symmetry. Central Nb atom (Nb1) labeled with schematic $d_{z^2}$ orbital to indicate spatial distribution. Scanning parameters: **a**, bias voltage $V_B = -1.5$ V, tunneling current $I_t = 20$ pA; **c** and **d**, $V_B = -1.5$ V, $I_t = 500$ pA.

To understand the detailed structure of the observed CDW order with the SOD, we constructed schematic atomic models of an SOD pattern (Fig. 1e) and a single SOD (Fig. 1f). In one single SOD, 13 Nb atoms were categorized as the central atom (Nb$_1$), six at the nearest sites (Nb$_2$), and the other six at the next nearest sites (Nb$_3$). As shown in Fig. 1f, the central Nb atom is superimposed by its $d_{z^2}$ orbital, whereas the other surrounding Nb atoms contract towards the central Nb$_1$ and give rise to the fluctuation in the top Se plane[21]. The 12 Se atoms in the SOD unit were categorized into two groups: six Se atoms in the topmost plane (labeled by brown circles), and the other six in the bottom plane (labeled by light brown circles). The six topmost Se atoms contributed more to STM imaging and resulted in triangular protrusions in the SOD unit, as shown in Fig. 1c.

**Reversed spatial contrast of UHB.** To further study the local electronic structures of different sites in the CDW pattern, we performed constant-height $dI/dV$ scanning tunneling spectroscopy (STS) along line AA' (as shown in Fig. 2a), which is the long diagonal axis of several CDW rhombus unit cells. The high-quality large ML-1T NbSe$_2$ islands enable us to select an area away from vacancy defects, adsorbates, or step edges; therefore, defect-induced and band-bending effects near the step edge were excluded. A zoomed-in image is shown in Fig. 2b. Owing to the triangular shapes of the top sites, two inequivalent hollow sites appeared and were named Hollow-1 and Hollow-2, as indicated by the blue triangle and green circle, respectively.

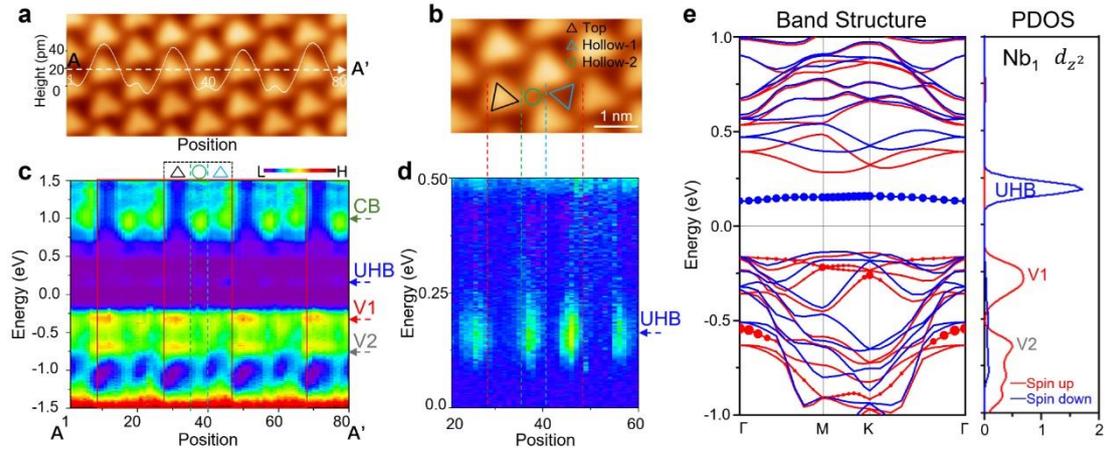

**Fig. 2 Spatial-dependent electronic structures of monolayer 1T-NbSe$_2$ on BLG/SiC(0001). a** Area of 1T-NbSe$_2$; AA' is the line along which line $dI/dV$ in **c** is obtained. Line profile (white curve) of AA' is shown above AA'. **b** Zoomed-in image of one CDW unit cell showing Top, Hollow-1, and Hollow-2 sites, labeled by black triangle, blue triangle, and green circle, respectively. **c** Line $dI/dV$ spectra along AA' shown in color map mode. Spectra in one unit cell is shown in the frame, which contains Top, Hollow-1, and Hollow-2 regions. Four orbitals are labeled as CB, UHB, V1, and V2. **d** Zoomed-in image of **c** in 0–0.5 eV range showing spatial dependence of UHB. **e** Left: calculated band structure of 1T-NbSe$_2$, where blue and red dots represent band projections of spin-down and spin-up dz$^2$ orbitals of the center Nb atom (i.e., Nb$_1$) in SOD, respectively. Right: corresponding partial density of states (PDOS) of Nb$_1$ dz$^2$ orbital. Calculations include spin polarization and GGA + U approximation. Scanning and spectroscopy parameters: **a** and **b** −1.5 V, 10 pA; **c** and **d** −1.5 V, 1 nA before turning off feedback loop.

The $dI/dV$ spectra along the AA' direction are shown in Fig. 2c and Supplementary Fig. 2, providing direct evidence of site-dependent change in the local electronic structure; it was observed that the change exhibited the periodicity of the CDW pattern. To provide visual guidance, we denoted four important orbitals (pronounced states at -0.28, -0.75, and +1.0 eV, and a faint state at +0.2 eV) by arrows, and named them as V1, V2, CB and UHB. ML-1T NbSe$_2$ is generally believed to be a Mott insulator[36,38,39], thus the states right above and below the Fermi level probably are the upper and lower Hubbard band (UHB and LHB), respectively. Our GGA+U calculation confirmed the Mott insulator nature and show the LHB is highly mixed with valence band (VB). Whereas energy change was not observed in these orbitals, the intensities of these orbitals changed significantly based on the CDW sites (e.g., along the AA' direction). For example, at the top sites, the V1 and V2 orbitals were strong, whereas the CB and UHB were weak; meanwhile, this trend was reversed at the hollow sites.

To distinguish the spatial distribution change in UHB state at +0.2 eV, we obtained high-precision line *dI/dV* spectra, which clearly show orbital feature with a periodicity (see Fig. 2d). Interestingly, the UHB exhibited a contrasting reversal with respect to the CDW topography (as denoted by the dashed-line in Figs. 2b and 2d). For example, at the bright top sites of the CDW superlattice, where the isolated $d_{z^2}$ orbitals of Nb atoms were located, the UHB was faint. Hence, our observations reveal that the Mott UHB was spatially distributed away from the $d_{z^2}$ orbital at the center of the CDW unit (the SOD).

To further investigate the characteristics of these four states, we conducted first-principles calculations on the band structure of ML-1T $NbSe_2$ under different conditions. At this stage, we considered the freestanding monolayer to obtain the pristine properties. Through a comparison of four different conditions (Supplementary Fig. 3), the following were observed: When both the spin polarization and GGA + U approximation were considered, a Mott gap size of ~0.42 eV was obtained with a Coulomb interaction U of 2.8 eV, as shown in Fig. 2e. The values of the Mott gap and U were similar with those reported in recent literature regarding $NbSe_2$[36,39,40]. Our combination of experimental *dI/dV* and theoretical study elucidated the characteristics of the faint state at +0.2 eV as the UHB. Furthermore, our calculation highlighted that the opening of the Mott gap was a result of the coupling between the Mott and magnetic effects. Notably, the UHB was isolated from the conduction band with a spin-down state that originated from the contribution of the $d_{z^2}$ orbital of the $Nb_1$ atom, whereas the corresponding LHB was mixed significantly with the valence band owing to the hybridization of the spin-up state with other orbitals.

After identifying these four important bands from the line *dI/dV* spectra, we further performed point spectroscopy and bias-dependent STM imaging (for 2D *dI/dV* mapping, see Supplementary Fig. 4) to provide a more comprehensive study regarding the spatial distribution of the electronic states. As shown in Fig. 3a, in the three curves obtained from the three typical sites (Top, Hollow-1, and Hollow-2), the main peak positions remained the same, whereas the peak intensities changed. To better describe this change, the top sites were labeled by yellow dots (as shown in Figs. 3b–3e) to signify the same or equivalent sites.

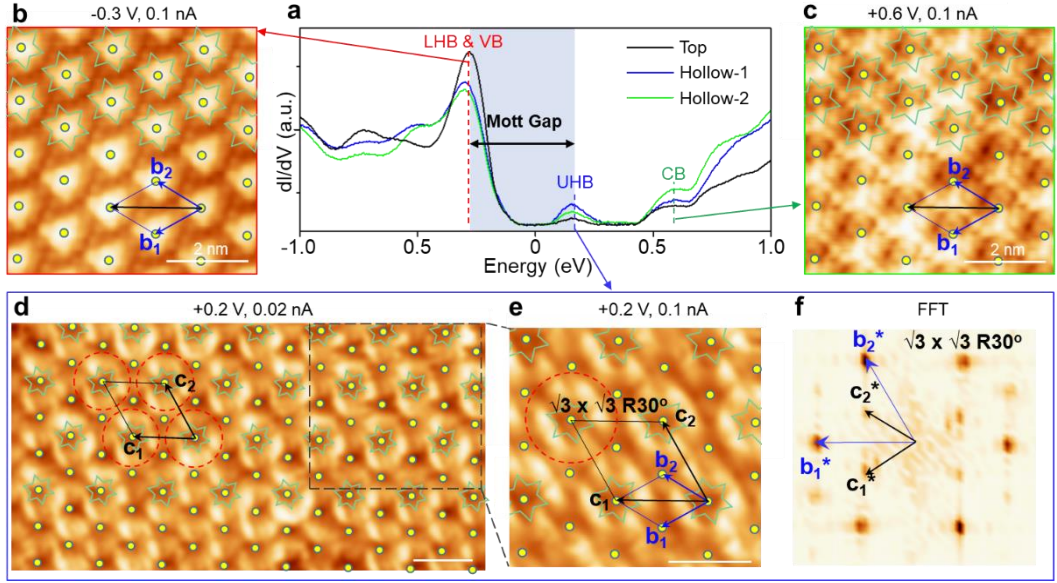

**Fig. 3 Intriguing spatial dependence of UHB. a** *dI/dV* spectra captured at Top, Hollow-1, and Hollow-2 areas of 1T-NbSe$_2$. **b, c** STM images of SL 1T-NbSe$_2$ at −0.3 and +0.6 V, respectively. **d, e** STM images captured with 0.2 V, 0.02 nA, and 0.2 V, 0.1 nA, respectively, showing √3 × √3 R30° superstructure (with unit vectors $c_1$ and $c_2$) with respect to the CDW pattern (with unit vectors $b_1$ and $b_2$). **f** Corresponding FFT of **e** tunneling parameters: **a** −1 V, 0.02 nA before turning off feedback loop. Scale bar for **b-e**: 2 nm.

**Extra superstructure of UHB.** The STM image captured at -0.3 eV (as shown in Fig. 3b) shows each triangular protrusion comprising six top Se atoms in each SOD area, resembling the image at -1.5 eV shown in Fig. 1c, and no clear change was visible. However, the STM image at +0.6 eV in Fig. 3c shows a clear change, i.e., the image contrast reversed: each top site marked by the yellow dot shows a clear change from a bright protrusion to a truncated triangular depression. This bias-dependent contrast is consistent with the black *dI/dV* spectrum shown in Fig. 3a, which shows that at -0.3 and +0.6 eV, the top exhibited the strongest and weakest intensities, respectively. This information is supported by the PDOS in Fig. 2e, which shows a very weak signal above UHB at the Nb$_1$ sites.

Subsequently, we focus on the interesting feature near the Fermi level: the state at +0.2 eV (UHB state). The image captured at a bias voltage of +0.2 eV shows that each top site is also a depression (Fig. 3d). This depression feature is consistent with the line *dI/dV* spectra shown in Fig. 2d. However, clear differences at +0.2 eV (Fig. 3d) and +0.6 eV (Fig. 3c) were observed. One difference was indicated by the red dashed-line circles in Fig. 3d, where the surrounding orbitals of the top site are not connected, but isolated as "bright petals of a flower".

Another difference was that a local super-period existed for the UHB states. Using the blue rhombus in Fig. 3e as an example, the two "flowers" centered at the long diagonal axis were brighter, whereas the two centered at the short diagonal axis were dim. Thus, in the hexagonal lattice of the SOD pattern, one bright flower is surrounded by six dim flowers as nearest neighbors. This spatial relationship resulted in a √3 × √3 R30° superstructure (with unit vectors $c_1$ and $c_2$) with respect to the CDW pattern (with unit vectors $b_1$ and $b_2$), as illustrated by the black rhombus in Fig. 3e. Subsequently, we further increased the feedback current from 0.1 to 3 nA to render the tip closer to the sample and resolve the atomic resolution, as shown in Supplementary Fig. 5, in which the atomic lattice in the center of the flower was resolved more effectively. The √3 × √3 R30° superstructure of the flower pattern can be well distinguished in the FFT pattern, as shown in Fig. 3f. Two sets of reciprocal lattices were resolved: the blue basis vectors represent the CDW lattice, and the black ones represent the √3 × √3 R30° with respect to the CDW pattern. Additional data of the √3 × √3 R30° superstructure were observed from a direct comparison of the STM images at -1.5 and +0.2 V, as shown in Supplementary Figs. 6, 7 and 8.

**CDW Pattern within the Mott gap.** After investigating the three pronounced states and the special UHB state of ML-1T NbSe$_2$, we further investigated the state around the Fermi level to identify possible gap states. Figure 4a shows a topographic image with six CDW long axis periods. Along the AA' direction, the line $dI/dV$ spectra in the range of -0.5–1.0 eV were obtained, as shown in Fig. 4b. We further zoomed-in on the energy region of -60–60 mV, which was close to the Fermi level but distant from the Hubbard band edge. The corresponding $dI/dV$ spectra are shown in Fig. 4c, in which some residual conductance is visible. The residual conductance (tunneling electrons) was reasonable as the STM tip was brought close to the sample surface and continued tunneling on the 1T-NbSe$_2$ even in the gap. The residual conductance (electronic state) was random and did not exhibit CDW patterns.

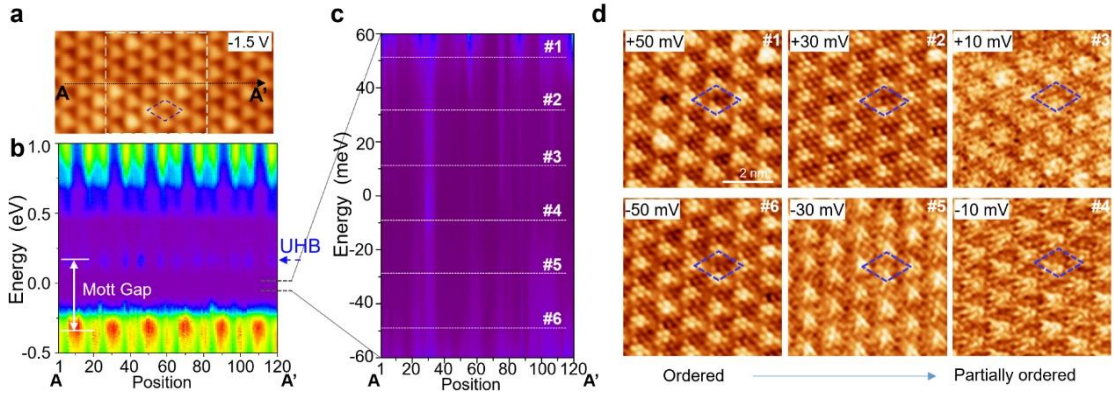

**Fig. 4 Resolving CDW pattern in Mott insulator gap. a** Area of 1T-NbSe$_2$; AA' is the line along which line $dI/dV$ spectra were obtained. **b** Line $dI/dV$ spectra captured along six CDW periods (AA' in **a**) showing Mott gap between UHB and LHB. **c** Zoomed-in line $dI/dV$ spectra in **b**, showing residual conductance within the gap, lacking periodicity. **d** Bias-dependent images showing CDW pattern clearly in Mott insulator gap. Energy positions are labeled in **c**, tunneling parameters: **a** −1.5 V, 10 pA; **b** −1.5 V, 1 nA; **c** −0.1 V, 0.5 nA before turning off feedback loop.

Intuitively, no CDW pattern is expected to be resolved in this energy region, and only atomic resolution is expected. However, both atom-resolved imaging and the CDW lattice were resolved clearly and ordered at ± 50 meV, and this phenomenon remained at ± 30 meV and persisted to ±10 meV, even though the CDW was only partially ordered. The ordered to partially ordered revolution suggested an electronic change among the images captured at different biases. In addition, the buckling of the top-hollow sites changed from 160 pm (at 50 mV) to 107 pm (at 10 mV), also indicating an electronic contribution change within the Mott gap. To the best of our knowledge, this is the first simultaneous observation of the residual conductance state in the Mott gap, bias-dependent CDW pattern, and atomic-resolved topmost lattice 2D TMD material in the same area.

**Discussion**

The spatial distribution of the Mott UHB at +0.2 eV is yet to be elucidated. In the literature of Density functional theory (DFT) calculations about $NbSe_2$[38,39,40], the UHB is believed to arise from the central Nb atom in the SOD with an isolated $d_{z^2}$ orbital, which is inconsistent with our observations that the electronic states at the top sites are relatively lower than the hollow sites (Fig. 2d and Fig. 3d). We then considered mainly orbitals of the periphery Nb atoms ($Nb_2$ and $Nb_3$ in Fig. 1f), and calculated the local density of states. As seen in Supplementary Fig. 9, the calculated results indicate that the observed electronic states are related with both the central Nb atom and the periphery Nb atoms. Interestingly, our room temperature (RT) and 77 K measurements reveal that STM images at both −0.3 eV and +0.2 eV show the same spatial distribution without any contrast reversal (Supplementary Fig. 10), in line with the Mottness behavior of $TaS_2$[21]. The unconventional UHB spatial distribution at low temperature is probably associated with a low temperature condensation process, similar to superconductivity, and paramagnetism. With the high CDW transition temperature above RT, our 1T-$NbSe_2$ facilitates the application of CDW pattern as a template for mono-atomic dispersion[41], such as heterogeneous catalysis, nanoscale magnetic quantum dots.

Whereas a Mott insulator with a UHB having a reversed contrast with the SOD top sites is interesting, the $\sqrt{3} \times \sqrt{3}$ R30° supermodulation at the UHB energy is even more intriguing. In this study, we excluded the substrate geometry effect such as the moiré pattern formed in the substrate. The BLG/SiC(0001) surface was reconstructed at the interface with a period of ~ 3 nm[42], and it did not exhibit a commensurate relationship with the $NbSe_2$ lattice of 1.25 nm. In addition, we studied $NbSe_2$ islands on BLG/SiC(0001) with different rotation angles (0° and 11°), and both showed the same supermodulation with a $\sqrt{3} \times \sqrt{3}$ R30° pattern (Supplementary Fig. 11). As our GGA + U calculations show that the UHB is spin-polarized, each SOD unit contains one spin at the center Nb atom (numbered as "1" in Fig. 1f), and the SOD lattice forms a triangular lattice. Hence, ML1T-$NbSe_2$ is very likely to be a platform to host antiferromagnets[43] or quantum spin liquid in a 2D triangular lattice[44-47] (Supplementary Fig. 12). Further investigations into this novel supermodulation are necessitated.

We note that there are some limitations with our current STM and DFT methods. The current STM measurements have not characterized the spin properties directly, and a combined measurement with a spin-polarized STM tip and external magnetic field would be very useful to provide direct evidence. For the DFT calculations, the used exchange-correlation functionals have not described the semi-local or long-range ordering precisely. Dynamic mean-field theory and density matrix renormalization group would be employed to give more information on the spin orderings. These limitations would be circumvented in future.

In conclusion, we comprehensively investigated the relationship between the Mott insulator and CDW in ML-1T NbSe$_2$ for the first time. Using STM/S to investigate the spatial and energy dependence of the correlated electronic system, we discovered that ML-1T NbSe$_2$ exhibited the Mott UHB feature away from the SOD center (top sites) in the CDW pattern. Moreover, the Mott UHB state demonstrated a local √3 × √3 R30° superstructure with respect to the typically observed CDW pattern. Furthermore, we observed a CDW pattern in the Mott gap, showing a correlation between the CDW and the Mott Hubbard band state. DFT calculations demonstrated that +0.2 eV was the spin-polarized UHB, whereas the LHB was mixed with the valence band. Using this quantum material, our findings provided new insights into the relationship between the correlated Hubbard band and CDW pattern in 2D systems.

**Methods**

**Sample preparation and STM measurements.** MBE and STM experiments were performed in an interconnected UHV system (base pressure 1 × 10$^{-10}$ mbar) comprising MBE chambers and a low-temperature STM chamber. Epitaxial graphene was grown by the thermal decomposition of 4H-SiC(0001) at 1200 °C for 40 min. NbSe$_2$ islands were prepared using the MBE method at a deposition rate of ~0.002 ML/min. An elemental Nb rod was used as the metal source in the e-beam evaporators (Focus Ltd.), whereas an elemental Se source in a Knudsen cell was heated to 120 °C. The deposition ratio of Se to Nb was larger than 20:1. During film deposition, the background pressure in the chamber was ~3.0 × 10$^{-9}$ mbar. The BLG/SiC substrate was stored at ~550 °C during the deposition and post-annealing process to improve the diffusion and desorb excess Se[48,49]. STM imaging was performed using mechanically cut PtIr tips at a liquid helium temperature (4.2 K). The *dI/dV* tunneling spectra were acquired using lock-in detection by applying an AC modulation of 20 mV (r.m.s.) at 973 Hz to the bias voltage. A typical *dI/dV* on BLG is shown in Supplementary Fig. 13.

**DFT calculations.** First-principles calculations were performed using the Vienna Ab Initio Simulation Package[50,51]. The projector-augmented wave method and generalized gradient approximation were used to describe the electronics[52,53]. The 4d and 5s electronics of Nb atoms were considered as valence electrons with a plane wave energy cutoff of 381 eV. A 5 × 5 × 1 Monkhorst–Pack[54] k-point mesh was adopted for geometric structure optimization, and a 7 × 7 × 1 k-point mesh was used to perform self-consistent calculations. To simulate the single layer situation, a 20 Å vacuum layer was applied. The relaxation of the atomic structure completed when the force on all atoms was < 0.01 eV/Å$^{-1}$.

**Data and materials availability**

All data needed to evaluate the conclusions in the paper are present in the paper and/or the Supplementary Materials. Additional data related to this paper are available from the corresponding author upon reasonable request.

**Acknowledgments**

Thanks for the financial supporting from National Key Research and Development Program of China (2019YFA0308000, 2020YFA0308800), National Natural Science Foundation of China (Nos. 61971035, 61725107, 61901038, 61922035, 11874171), Beijing Natural Science Foundation (Nos. Z190006, 4192054), and the Strategic Priority Research Program of Chinese Academy of Sciences (XDB30000000).

**Contributions**

L.W.L., H.J.G., and Y.L.W. coordinated the research project. L.W.L. performed the main experimental fabrication and measurements. H.Y., X.S., Q.Z.Z., Z.P.H., Y.H.H., Y.Y.C., Z.Q.X., T.Z., X.W., Y.H., and Y.L.W. partially participated in the sample fabrication and measurements. Y.T.H., J.T.S., X.B.L., F.W.Z. and Y.G.Y. performed the DFT calculations. All authors analyzed the data and discussed the manuscript.

**Additional information**

**Competing interests**

The authors declare no competing interests.